 \documentclass[final,3p,times,onecolumn]{elsarticle}

 \usepackage{graphics}
 \usepackage{graphicx}

\usepackage{amssymb}
 \usepackage{amsmath}

\journal{Journal of Computational Physics}

\begin{document}

\begin{frontmatter}



\title{On the velocity space discretization for the Vlasov-Poisson system: comparison between Hermite spectral and Particle-in-Cell methods.\\
Part 1: semi-implicit scheme}


\author[1]{E. Camporeale}
\author[1]{G.~L. Delzanno}
\author[2]{B.~K. Bergen}
\author[1]{J.~D. Moulton}

\address[1]{T-5 Applied Mathematics and Plasma Physics, Los Alamos National Laboratory, 87545 Los Alamos, NM, USA.}
\address[2]{CCS-7 Applied Computer Science, Los Alamos National Laboratory, 87545 Los Alamos, NM, USA.}
\begin{abstract}
We discuss a spectral method for the numerical solution of the Vlasov-Poisson system where the velocity space is decomposed by means of an Hermite basis.
We describe a semi-implicit time discretization that extends the range of numerical stability relative to an explicit scheme.
We also introduce and discuss the effects of an artificial collisional operator, which is necessary to take care of the velocity space filamentation problem, unavoidable
in collisionless plasmas.
The computational efficiency and the cost-effectiveness of this method are compared to a Particle-in-Cell (PIC) method in the case of a two-dimensional phase space.
The following examples are discussed: Langmuir wave, Landau damping, ion-acoustic wave, two-stream instability, and plasma echo.
The Hermite spectral method can achieve solutions that are several orders of magnitude more accurate (at a fraction of the cost) with respect to the PIC method.
\end{abstract}


\end{frontmatter}


\section{Introduction}\label{Intro}
The Vlasov equation represents the cornerstone for the kinetic modeling
of collisionless plasmas. It describes the time evolution of the distribution function of a population of charged particles that respond self-consistently to the effect of self- and externally 
induced electromagnetic fields.
The numerical solution of the Vlasov equation for collisionless plasmas represents a very active area of research.
The main challenge resides in the fact that the distribution function lives in a six dimensional phase space.
Unarguably, the most widely used technique to solve the Vlasov equation is the Particle-In-Cell (PIC) method \citep{birdsall_book,hockney_book}.
The main idea, originally developed in the 50's, is to sample the distribution function in velocity space by means of a discrete number of super-particles.
The electromagnetic field is represented on a grid in the computational domain, and the super-particles move through the grid according to the Lorentz force that is calculated by interpolation 
from the grid to the particles position.
The particles interactions are therefore mediated by the grid, and in this way the number of operations per time step is reduced from $\sim N_p^2$ (if the total force on a particle is calculated
by summing the pair interaction with every other particle) to $\sim N_p$ (with $N_p$ the number of super-particles).
A comprehensive review of the PIC methodology can be found in \citep{verboncoeur05}.

The primary shortcoming of PIC simulations is related to the numerical noise: even starting from an equilibrium configuration, the discrete nature of the particles generates instantaneous
(i.e. within the first time step) unphysical perturbations that produce a 'noise ground' level in the fields, below which any physical signal is lost.
The most obvious way to reduce the noise in PIC is by increasing the number of super-particles, i.e. refining the discretization in velocity space.
The main problem is that while the simulation time scales roughly linearly with $N_p$, the noise ground level decreases only as $N_p^{-1/2}$, as typical of Monte Carlo methods, implying that problems 
that require a high signal-to-noise ratio require significant computational resources.
Some examples that illustrate the impact of the particle noise on the efficiency of PIC codes are the recent studies in space plasmas
on the acceleration of particles in the solar wind and the coupling between large scale
turbulence and small scale (i.e. kinetic) dissipative effects \citep{valentini08,matteini12, haynes13}.
PIC codes have demonstrated very poor performances for these problems. In two-dimensional (2D)
simulations, \citet{camporeale11} have shown that at least $\sim 10,000$ particles per cell (equivalent
to $\sim10^6$ in 3D) are needed to achieve a sufficiently large signal-to-noise ratio. For comparison, one of the biggest and most advanced PIC simulations in the world (of a magnetic reconnection problem) 
was done with ~250 particles per cell
and required the use of the highest performance, petascale computer available at that time \citep{daughton11,bowers09}.
We note that methods based on $\delta$f or re-mapping of the distribution function have been developed in the PIC community to reduce particle noise. This has led to some very interesting work (see for instance \citep{chehab05, wang11, deng14}), but has not yet resulted in a commonly accepted denoising technique adopted by the PIC community.

The non-optimal scaling of the noise level with the number of particles in the PIC method has long been recognized in the computational plasma physics community \citep{byers70, denavit81, sydora99, nevins05}, 
and it has perhaps been the main motivation for developing alternative ways of solving the Vlasov equation numerically.
Nevertheless, despite its intrinsic noise, the PIC method still represents the preferred approach in the community, probably due to its simplicity of coding, the relative ease to parallelize it, 
and the recent impressive advances in available computing power. However, one should always keep in mind the resource-intensive nature of PIC and the related cost. As a figure of merit on the cost 
of large scale PIC simulations, a single run using the global gyrokinetic PIC code developed within the SciDAC Gyrokinetic Particle Simulation Center (GPSC) \citep{batchelor07} for investigating 
the long-time evolution of turbulent transport in nuclear fusion devices was estimated, in 2008, to take approximately 25 million CPU hours, on 100,000+ cores \citep{tang08}. The monetary cost of 
such a simulation can be estimated at about \$1,000,000 \citep{walker09}.
In view of the upcoming exascale era and the complexity of multiscale plasma physics, it is therefore legitimate to question the computational efficiency of the PIC algorithm and wonder whether 
it is possible to devise algorithms that use the available computational resources more efficiently.

An alternative to the PIC method is represented by solving the Vlasov equation directly in phase-space (namely with a sixth dimensional computational grid in space and velocity),
by means of a so-called Eulerian Vlasov code.
This has been done, usually for problems with reduced dimensionality, with a variety of techniques such as high-order finite differences \citep{whealton86, guo12}, finite elements \citep{besse03,heath12},
or finite volumes \citep{banks10,duclous12}.
The reader is referred to \citep{shoucri08} for a recent review of different methods.
Notably, successful algorithms include the semi-Lagrangian schemes \citep{shoucri74,cheng76,sonnendrucker99,besse03,umeda06,carrillo07,crouseilles07,imadera09,crouseilles10,qiu10,qiu11, rossmanith11}
and positive flux conservative methods \citep{filbet01}.\\
Yet another class of techniques is represented by spectral methods where the velocity space is projected onto a complete orthogonal basis,
and therefore the phase space is effectively not gridded in the velocity coordinates \citep{engelmann63, armstrong70, denavit71, klimas83, eliasson03, lebourdiec06}.\\
Clearly, all these methods have advantages and shortcomings. Historically, the PIC method has been favored for large-scale, multidimensional problems which are still
not doable with Vlasov solvers (mainly due to memory limitations) \citep{mangeney02}. On the other hand, Vlasov codes allow the study of fine scale details of the distribution functions that are typically inaccessible
in PIC codes, due to the above mentioned noise problem.
Vlasov codes, however, can suffer from serious numerical problems that can lead to a lack of discrete conservation properties, the occurrence of unphysical oscillations, and
the generation of non-positive values in the distribution functions \citep{banks10}.\\
In this paper, we focus on a spectral method where the distribution function is projected onto an Hermite basis in velocity space.
This gives rise to a set of nonlinear, time- and space-dependent PDEs for the coefficients of the expansion.
The use of Hermite functions to discretize the velocity space in the Vlasov equation dates back to the works of \citet{grad49, engelmann63, grant67a}, and \citet{armstrong70}.
More recently this approach has been investigated by \citet{holloway96,schumer98,lebourdiec06} and, in the context of linearized equations, by \citet{camporeale06} and \citet{siminos11}.

The expansion of the distribution function using an Hermite basis in velocity space can be appealing for several reasons.
First, the Hermite functions are a complete orthogonal basis with respect to a Gaussian
weight function. As such, they are the optimal basis to represent Maxwellian-like distributions. In fact, an exact Maxwellian in velocity can be represented by only one term of the Hermite basis.
Second, as already noted by \citet{grad49}, the low order expansion coefficients are directly related to the low order moments of the distribution functions
(i.e. density, mean velocity, energy, heat flux, etc.) that describe the plasma macroscopically and are usually the physical quantities of interest.
As a consequence, the use of the Hermite basis allows a smooth transition from a fluid to a kinetic description by simply increasing the number of coefficients retained.
This is an important and crucial feature of this method in approaching multi-scale problems and in assessing the importance of kinetic effects, which is not available in PIC or Vlasov methods that are 
forced to treat the full distribution function.

This article describes a semi-implicit discretization in time of the truncated set of PDEs for the expansion coefficients.
A fully-implicit time discretization is discussed in the companion paper (Part 2) \citep{camporeale13b}.
The spatial dependence of the coefficients is represented by means of a discrete Fourier transform. \\
This paper builds largely upon the work of \citet{schumer98}.
In \citep{schumer98} two different Hermite bases (so called 'asymmetrically' and 'symmetrically' weighted) and their properties were discussed, and a qualitative comparison between the Fourier-Hermite (FH)
method and the PIC method was presented for simulations of Landau damping and bump-on-tail instability. Here we expand that work in three ways.
First, we propose a semi-implicit scheme that is numerically more stable than the explicit scheme presented in \citep{schumer98}.
Second, we study the effects of an artificial collision operator, with the intent of controlling the filamentation process and suppressing/mitigating numerical instabilities.
Third, we \textit{quantitatively} compare the FH method against a conventional explicit PIC code.
In order to assess which method must be preferred (and for which conditions), the key information is represented by the CPU time needed to obtain a solution with a certain accuracy
(this metric was not considered in \citet{schumer98}).
Hence, our comparisons are presented in terms of computational efficiency and efficacy. The former is essentially represented by the CPU time required to achieve a certain accuracy in the solution, 
while the latter is a measure
of the cost-effectiveness of the algorithm, i.e. how much an increased resolution and consequently a longer CPU time is paid off in terms of better accuracy \citep{lapenta06}.\\
Although it is not surprising that a spectral method performs better than the super-particle discretization used in PIC, we are not aware of any other study that
presents such quantitative comparison in terms of computational efficiency and efficacy.
For the examples presented in this paper (involving mostly near-equilibrium Maxwellian plasmas in one dimension), the difference in performance between the FH method and the PIC is impressive, with the former
producing results that are several orders of magnitude more accurate for equal CPU time or, conversely, results with the same level of accuracy in a fraction of CPU time. Nevertheless, FH is still 
in its 'infancy' and the development of optimization techniques that can accelerate the convergence of the Hermite basis will be critical for multidimensional applications, particularly for plasmas 
away from equilibrium.

The paper is organized as follows. Section 2 presents the mathematical foundation of the FH method, along with its conservation properties.
Section 3 discusses the time discretization of the equations. Section 4 is devoted to a brief discussion of the filamentation problem in Vlasov codes, and how that translates to PIC codes.
It also introduces the collisional term that we have implemented and used in some of the simulations. Section 5 presents the comparison between PIC and FH methods
for five classical test cases: Langmuir wave, Landau damping, ion-acoustic wave, two-stream instability, and plasma echo. Finally, our conclusions are presented in Section 6.

\section{Numerical Method: Fourier-Hermite basis}\label{hermite}
We study the Vlasov-Poisson system in the 2-dimensional phase space $(x,v)$, where $x$ denotes position and $v$ velocity.  
The phase space is assumed to be periodic in $x$.
In order to describe the method we specialize to the case of a plasma consisting of an electron and a singly charged ion species. 
The quantities are normalized as follows. Time is normalized on the electron plasma frequency $\omega_{pe}$, velocities on
the electron thermal velocity $v_{te}=\sqrt{kT_e/m_e}$, lengths on the electron Debye length $\lambda_D$, the electric field on $\frac{m_e}{e}v_{te}\omega_{pe}$ ($k$ is the Boltzmann's constant, $T_e$ is 
the electron temperature,
$m_e$ is the electron mass and $e$
is the elementary charge), and densities on a reference density $n_0$.
The Vlasov equation for the species $s$ reads:
\begin{equation}
 \frac{\partial f_s}{\partial t} + v\frac{\partial f_s}{\partial x} + \frac{q_sm_e}{em_s} E\frac{\partial f_s}{\partial v}=0.\label{vlasov}
\end{equation}
Here $f_s$ is the particle distribution function, $q_s$ and $m_s$ are the charge and mass of the particles of species $s$ ($s=e,i$ for electrons and ions, respectively) and $E$ is the electric field.
The electric field is self-consistently calculated through the Poisson equation:
\begin{equation}
\frac{\partial E}{\partial x}= \int^\infty_{-\infty} f_i dv-\int^\infty_{-\infty} f_e dv.\label{poisson}
\end{equation}
The discretization in velocity employs the asymmetrically-weighted Hermite basis:
\begin{eqnarray}
\Psi_n(\xi)&=&(\pi2^nn!)^{-1/2}H_n(\xi)e^{-\xi^2}\label{H_basis_1}\\
\Psi^n(\xi)&=&(2^nn!)^{-1/2}H_n(\xi)\label{H_basis_2}
\end{eqnarray}
where $H_n$ is the $n$-th Hermite polynomial.
The distribution function $f(x,v,t)$ is defined as:
\begin{equation}
 f_s(x,v,t)=\sum_{n=0}^{N_H-1} C^s_n(x,t) \Psi_n(\xi_s)\label{f},
 \end{equation}
with $\xi_s=(v-u_s)/\alpha_s$, where $u_s$ and $\alpha_s$ are two constant parameters for each species, and $N_H$ is the number of Hermite modes (equal for both species).
The following closure is used for both species: $C_n^s=0$ for $n\geq N_H$.
The spatial dependence of the expansion coefficients $C^s_n$ will be treated later by means of a Fourier decomposition.
We note that the completeness of the Hermite basis  does not depend on the choice of the free parameters $u_s$ and $\alpha_s$.
It is well known, however, that a proper choice of the rescaling coefficient $\alpha_s$ can substantially accelerate the convergence of the series \citep{schumer98,tang93, camporeale06}. 
In this work, we allow even more
flexibility by introducing the free parameter $u_s$, which is a shift in the Hermite function argument. The usefulness of such a parameter will be discussed in Section \ref{twostream}.
Formulas for calculating the optimal values for $\alpha_s$ and $u_s$ were presented in \citet{camporeale06}.\\
The Hermite basis has the following properties ($\delta_{n,m}$ is the Kronecker delta):
\begin{eqnarray}
\int^\infty_{-\infty} \Psi_n(\xi)\Psi^m(\xi) d\xi = \delta_{n,m},\\
v\Psi_n(\xi) = \alpha\sqrt{\frac{n+1}{2}}\Psi_{n+1}(\xi)+\alpha\sqrt{\frac{n}{2}}\Psi_{n-1}+u\Psi_n,\\
\frac{d\Psi_n(\xi)}{dv} = -\frac{1}{\alpha}\sqrt{2(n+1)}\Psi_{n+1}(\xi).
\end{eqnarray}
One can multiply Eq. (\ref{vlasov}) by $\Psi^n(\xi)$, integrate in $d\xi$ and, by using such properties, obtain:
\begin{equation}\label{main_eq}
 \frac{\partial C^s_n}{\partial t}  + \alpha_s\left(\sqrt{\frac{n+1}{2}}\frac{\partial C^s_{n+1}}{\partial x} + \sqrt{\frac{n}{2}}\frac{\partial C^s_{n-1}}{\partial x}+\frac{u_s}{\alpha_s}\frac{\partial C^s_n}{\partial x}\right) -\frac{q_sm_e}{em_s}\frac{\sqrt{2n}}{\alpha_s}E C^s_{n-1}=0.
\end{equation}
Similarly, the Poisson equation (\ref{poisson}) becomes:
\begin{equation}\label{poisson_h}
 \frac{\partial E(x)}{\partial x}=\alpha_i C^i_0(x)-\alpha_e C^e_0(x).
\end{equation}
One can notice the well-known property that the electric field is related uniquely to the 0-th order expansion coefficients $C^s_0$, i.e., the density.
By now treating $E(x)$ and $C_n(x,t)$ by means of a standard discrete Fourier decomposition (with $L$ the domain length, and $2N+1$ modes):
\begin{eqnarray}
 C^s_n(x) &=& \sum_{k=-N}^N C^s_{n,k} e^{\frac{2\pi ikx}{L}}\\
 E(x) &=& \sum_{k=-N}^N E_k e^{\frac{2\pi ikx}{L}},\\
\end{eqnarray}
and using the orthogonality of the Fourier basis, one can finally derive an infinite set of ordinary differential equations for the coefficients $C^s_{n,k}(t)$:
\begin{equation}\label{vlasov_hf}
 \frac{dC^s_{n ,k}}{d t}  + \alpha_s\frac{2\pi ik}{L}\left(\sqrt{\frac{n+1}{2}}C^s_{n+1,k} + \sqrt{\frac{n}{2}} C^s_{n-1,k}+\frac{u_s}{\alpha_s}C^s_{n,k}\right) -\frac{q_sm_e}{em_s}
 \frac{\sqrt{2n}}{\alpha_s}\sum_{m=-N}^N E_{k-m} C_{n-1,m}^s=0,
\end{equation}
where $n$ $(k)$ is the index associated with the Hermite (Fourier) basis.
The expression for the electric field reads:
\begin{eqnarray}\label{poisson_hf}
  E_k &=& \frac{iL}{2\pi k}\left(\alpha_eC_{0,k}^e-\alpha_i C_{0,k}^i\right)   \mbox{       for $k\neq0$,} \\
  E_0 &=& 0.
\end{eqnarray}
We note that the Fourier representation of Poisson equation (\ref{poisson_h}) leaves the constant $E_0$ undefined, while imposing the constraint $\alpha_iC^i_{0,0}=\alpha_eC^e_{0,0}$
(which physically means that the plasma is neutral). 
However, the absence of an externally imposed electric field and the periodicity of the domain dictates that $E_0=0$.\\
The solution of Equation (\ref{vlasov_hf}) is the main objective of this paper. 
An important point to realize is that the only non-linearity of this equation is in the last term, which couples the electric field with the distribution function via a convolution.
Also, in the Hermite (velocity) space, the $n$-th coefficient $C_n$ is coupled only to $C_{n-1}$, $C_{n+1}$, to itself (if $u_s\neq 0$), and to $C_0$.
\subsection{Conservation properties}
As noted in \citet{schumer98}, the decomposition in the asymmetric Fourier-Hermite basis allows conservation of particle number and momentum exactly, irrespective of the chosen
time discretization scheme. On the other hand, exact energy conservation is not possible with the operator splitting scheme investigated in this paper. 
It is however possible to recover exact energy conservation by employing a fully-implicit time discretization, as discussed in the 
companion paper (Part 2) \citep{camporeale13b}.\\
Following \citep{schumer98}, the total number of particles for each species is defined as
\begin{equation}
 M_s=\int_0^L\int_{-\infty}^\infty f_s(x,v,t) dvdx = L\alpha_s C_{0,0}^s.
 \end{equation}
By inspection of Eq. (\ref{vlasov_hf}) one can see that $\frac{dC_{0,0}^s}{d t}=0$, from which the conservation of mass follows.\\
The single species linear momentum is defined as:
\begin{equation}
 P_s=\frac{m_s}{m_e}\int_0^L\int_{-\infty}^\infty vf_s(x,v,t) dvdx = \frac{m_s}{m_e}\left(\frac{L\alpha_s^2}{\sqrt{2}} C_{1,0}^s + u_s\alpha_s L C_{0,0}^s\right).
 \end{equation}
In order to prove the conservation of momentum, one can calculate the time derivative of the coefficients $C_{1,0}^i$ and $C_{1,0}^e$:
 \begin{eqnarray}
 \frac{dP_i}{dt} &=& \frac{m_i}{m_e}\frac{L\alpha_i^2}{\sqrt{2}} \frac{dC_{1,0}^i}{dt} = L\alpha_i\sum_{m=-N}^NE_{-m}C_{0,m}^i\\
 \frac{dP_e}{dt} &=& \frac{L\alpha_e^2}{\sqrt{2}} \frac{dC_{1,0}^e}{dt} = -L\alpha_e\sum_{m=-N}^NE_{-m}C_{0,m}^e.\label{Pe}
\end{eqnarray}
One can use Eq. (\ref{poisson_hf}) to substitute $C_{0,m}^e$ in Eq. (\ref{Pe}) and obtain:
\begin{equation}\label{P_cons}
 \frac{dP_e}{dt} = -L\alpha_e\sum_{m=-N}^N  E_{-m}\left(\frac{\alpha_i}{\alpha_e}C_{0,m}^i-\frac{2\pi mi}{L\alpha_e}E_m\right)=-L\sum_{m=-N}^N\left(  \alpha_iE_{-m}C_{0,m}^i-\frac{2\pi mi}{L}|E_m|^2\right).
\end{equation}
By taking into account that $E_{-m}$ is the complex conjugate of $E_m$ (since the electric field in physical space is a real quantity), 
it follows that $|E_{-m}|^2=|E_m|^2$ and therefore the last term in parenthesis in Eq.(\ref{P_cons}) is zero. Hence $\frac{dP_i}{dt} +\frac{dP_e}{dt}$=0.\\

\section {Time discretization}
In the literature it has been suggested that the Vlasov-Poisson system can be efficiently solved by means of an operator splitting procedure \citep{cheng76}, 
i.e., by decoupling the advection and the acceleration terms,
which are usually referred to as 'X-shift' and 'V-shift' operators, respectively.
Therefore, we define the $\mathbf{X}$ and the $\mathbf{V}$ operators acting on $C^s_{n,k}$ as: 
\begin{eqnarray*}
 \mathbf{X}[C^s_{n,k}] &=& - \alpha_s\frac{2\pi ik}{L}\left(\sqrt{\frac{n+1}{2}}C^s_{n+1,k} + \sqrt{\frac{n}{2}} C^s_{n-1,k}+\frac{u_s}{\alpha_s}C^s_{n,k}\right)\\
 \mathbf{V}[C^s_{n,k}] &=& \frac{q_sm_e}{em_s}
 \frac{\sqrt{2n}}{\alpha_s}\sum_{m=-N}^N E_{k-m} C_{n-1,m}.
\end{eqnarray*}
By following the second-order splitting scheme introduced by \citet{strang68}, the Vlasov equation is solved by updating the distribution function from time $t_0$ to time $t_0+\Delta t$
in the following three steps (subscript $s$ is omitted):\\
\begin{enumerate}
 \item  Solve $\frac{dC_{n,k}(t)}{dt} = \mathbf{X}[C_{n,k}(t)]$, evolving $C_{n,k}(t_0)$ into $C^*_{n,k}$ by an half time step $\Delta t/2$. Calculate the electric field from $C^*_{n,k}$;\\ 
 \item  Solve $\frac{dC_{n,k}(t)}{dt} = \mathbf{V}[C_{n,k}(t)]$, evolving $C^*_{n,k}$ into $C^{**}_{n,k}$ by a full time step $\Delta t$, using the electric field calculated in 1);\\
 \item  Solve $\frac{dC_{n,k}(t)}{dt} = \mathbf{X}[C_{n,k}(t)]$, evolving $C^{**}_{n,k}$ into $C_{n,k}(t_0+\Delta t)$ by an half time step $\Delta t/2$.\\ 
\end{enumerate}

Reference \citep{schumer98} has employed this scheme by using an explicit fourth-order Runge-Kutta (RK4) integration for both the X- and V- shifts.
They justify this choice by arguing that it offers a reasonable trade between accuracy and stability. However, we have verified that such a scheme has a very limited stability region,
even with relatively small time step.
In this paper, we improve on the stability of the explicit scheme by introducing a semi-implicit scheme. The V shift is still treated with RK4, but the X-shift is treated implicitly.
We have tested both a first order implicit Euler scheme (BDF1) \citep{henrici62} and a second-order Cranck-Nicolson \citep{crank47} 
scheme for the X-shift. 
Note that the Strang splitting ensures the second order accuracy in time of the overall scheme.
In summary, the discretized form of the X-shift that advances the solution
for half time step is the following (where the superscript $t$ indicates the time step):\\
\subsection*{X-shift (BDF1)}
\begin{equation}
  \left(1 + \Delta t u\frac{\pi ik}{L} \right)C_{n ,k}^{t*} + \Delta t\alpha\frac{\sqrt{2}\pi ik}{2L}\left(\sqrt{n+1}C_{n+1,k}^{t*} + \sqrt{n} C_{n-1,k}^{t*}\right) =  C_{n,k}^{t_0}
\end{equation}
\subsection*{X-shift (CN)}
\begin{align}
  \left(1 + \Delta t u\frac{\pi ik}{2L} \right)C_{n ,k}^{t*} + \Delta t\alpha\frac{\sqrt{2}\pi ik}{4L}\left(\sqrt{n+1}C_{n+1,k}^{t*} + \sqrt{n} C_{n-1,k}^{t*}\right) =\\
  =\left(1 - \Delta t u\frac{\pi ik}{2L} \right)C_{n ,k}^{t_0} - \Delta t\alpha\frac{\sqrt{2}\pi ik}{4L}\left(\sqrt{n+1}C_{n+1,k}^{t_0} + \sqrt{n} C_{n-1,k}^{t_0}\right)
\end{align}
Of course, both the BDF1 and the CN schemes can be written in matrix form. For each Fourier component $k$, we have:
\begin{equation}
 \mathbf{X}_{\frac{\Delta t}{2}}^k C_{k}^{t*} = \mathbf{B}_{\frac{\Delta t}{2}}^k C_{k}^{t_0}
\end{equation}
where $C_{k}$ is the column-vector $(C_{0,k}, C_{1,k}, C_{2,k},\ldots)^T$.
The matrix $\mathbf{X}$ is defined as
\begin{align}
 {X}_{\Delta t}^k (m,m) &= 1 + \theta\Delta t u\frac{\pi ik}{L}\\
 {X}_{\Delta t}^k (m,m-1) &= \theta\Delta t\alpha\frac{\sqrt{2(m-1)}\pi ik}{2L}\\
 {X}_{\Delta t}^k (m,m+1) &= \theta\Delta t\alpha\frac{\sqrt{2m}\pi ik}{2L}
\end{align}
with $\theta=1$ for CN and $\theta=2$ for BDF1. 
For the BDF1 scheme the matrix $\mathbf{B}$ is the identity matrix, and for the CN scheme is defined as:
\begin{align}
 {B}_{\Delta t}^k (m,m) &= 1 - \Delta t u\frac{2\pi ik}{L}\\
 {B}_{\Delta t}^k (m,m-1) &= -\Delta t\alpha\frac{\sqrt{2(m-1)}\pi ik}{2L}\\
 {B}_{\Delta t}^k (m,m+1) &= -\Delta t\alpha\frac{\sqrt{2m}\pi ik}{2L}.\\
\end{align}
Note that the matrices $\mathbf{X}$ and $\mathbf{B}$ commute in both cases (BDF1 and CN). Therefore the two consecutive X-shifts in 
steps 3) and 1): $C^{**}\overset{\Delta t/2}{\rightarrow}C^{t+\Delta t}\overset{\Delta t/2}{\rightarrow}C^{*}$ can be computed exactly as a single step:
\begin{equation}
\mathbf{X}_{\Delta t/2}\mathbf{X}_{\Delta t/2}C^*_{k} = \mathbf{X}_{\Delta t/2}\mathbf{B}_{\Delta t/2}C_{k}^{t+\Delta t} = \mathbf{B}_{\Delta t/2}\mathbf{X}_{\Delta t/2}C_{k}^{t+\Delta t}= \mathbf{B}_{\Delta t/2}\mathbf{B}_{\Delta t/2}C^{**}_{k}
\end{equation}
that is
\begin{equation}\label{Xshift}
\left(\mathbf{X}_{\Delta t/2}\right)^2C_{k}^* = \left( \mathbf{B}_{\Delta t/2}\right)^2C_{k}^{**}.
\end{equation}
We solve Eq. (\ref{Xshift}) by performing an LU decomposition of the matrix $\left(\mathbf{X}_{\Delta t/2}\right)^2$.\\

The V-shift is a standard fourth order Runge-Kutta integrator, reported here for completeness.
\subsection*{V-shift (RK4)}
\begin{figure}
\centering
\includegraphics[width=0.8\textwidth]{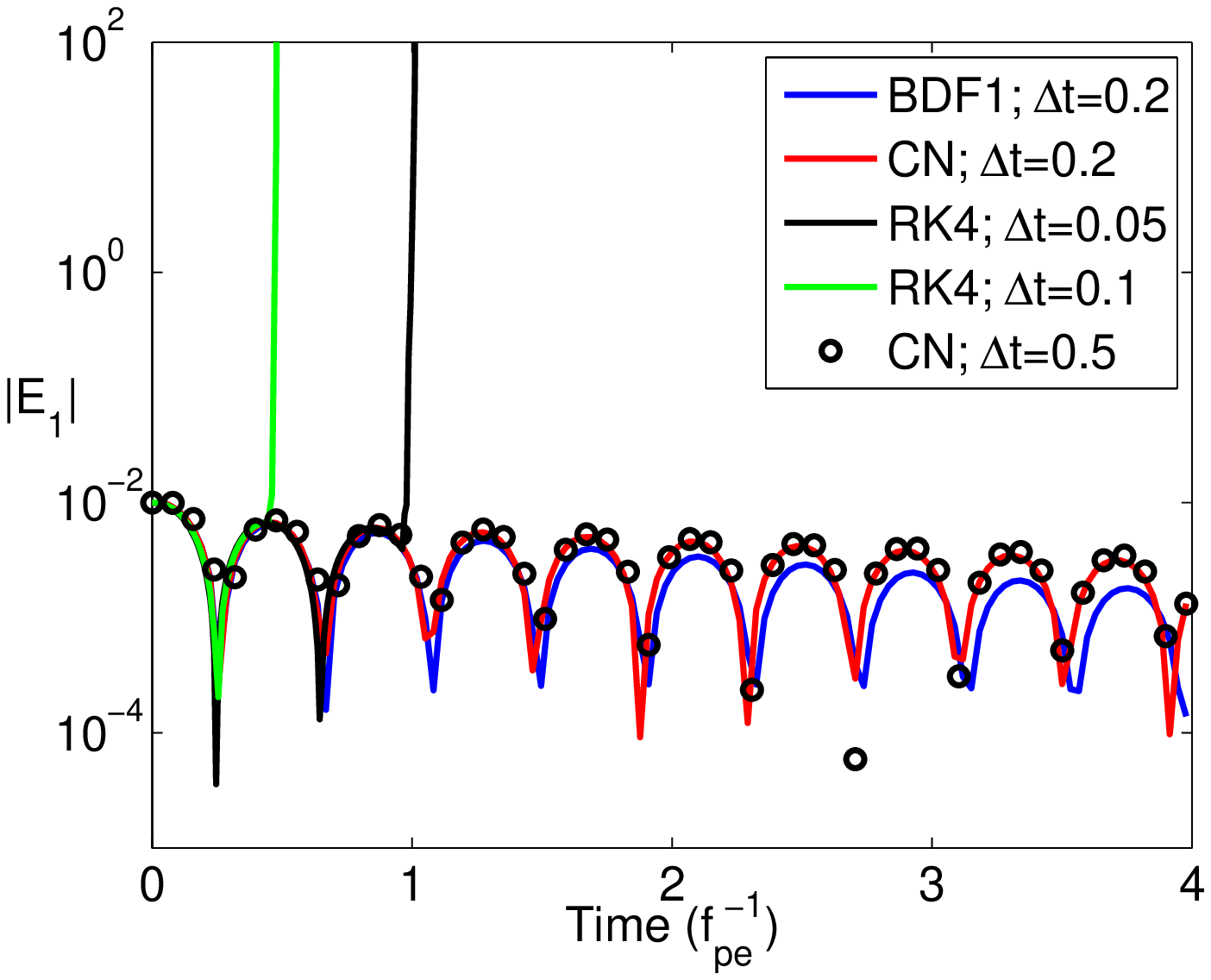}
\caption{Landau damping simulation with parameters: $\alpha_e=1$, $\varepsilon=0.01$ The X-shift is solved with BDF1 (blue line), CN (red line and black circles), and RK4 (black and green line). BDF1 and CN 
result in higher stability: the solution is stable for time step $\Delta t =0.5$, while the RK4 solution is unstable for time step as small as $\Delta t=0.05$.}\label{fig:stability}
\end{figure}

\begin{eqnarray*}
 h^1_{n,k} &=&-\Delta t\frac{q_sm_e}{em_s} \frac{\sqrt{2n}}{\alpha_s}\sum_{m=-N}^N E_{k-m} C_{n-1,m}^*\\
 h^2_{n,k} &=&-\Delta t\frac{q_sm_e}{em_s} \frac{\sqrt{2n}}{\alpha_s}\sum_{m=-N}^N E_{k-m} \left(C_{n-1,m}^*+\frac{1}{2}h^1_{n,k}\right)\\
 h^3_{n,k} &=&-\Delta t\frac{q_sm_e}{em_s} \frac{\sqrt{2n}}{\alpha_s}\sum_{m=-N}^N E_{k-m} \left(C_{n-1,m}^*+\frac{1}{2}h^2_{n,k}\right)\\
 h^4_{n,k} &=&-\Delta t\frac{q_sm_e}{em_s} \frac{\sqrt{2n}}{\alpha_s}\sum_{m=-N}^N E_{k-m} \left(C_{n-1,m}^*+h^3_{n,k}\right)\\
 C_{n,k}^{**} &=& C_{n,k}^* +\frac{1}{6}(h^1_{n,k}+h^4_{n,k}) + \frac{1}{3}(h^2_{n,k}+h^3_{n,k}).
\end{eqnarray*}
Note that the V-shift calculated via RK4 requires the evaluation of four convolutions for each Hermite mode $n=1,\ldots,N_H-1$.
Convolution is an expensive operation, which has a cost $\sim N^2$ if performed explicitly. A cheaper alternative is to
exploit the property that the Fast Fourier Transform (FFT) of the convolution of two vectors is equal to the product of the FFT of the two vectors.
In this way, the cost is reduced to $\sim N\log N$ (see, e.g. \citep{bracewell_book}). 
In practice, this consists in transforming the electric field and the coefficients $C_n$ from Fourier to physical space, and transforming their product back from physical to Fourier space at each time step.\\
In Figure \ref{fig:stability}, we present an empirical proof of the enhanced stability of the semi-implicit scheme (X-shift treated with BDF1 or CN) with respect to the fully explicit scheme 
(X-shift treated with RK4). This example is for the study of Landau damping, discussed in more detail in the next Sections. The blue and red lines show the evolution of the electric field
using respectively a BDF1 and a CN scheme, with time step $\Delta t=0.2$. The BDF1 shows its mildly dissipative character by having a stronger damping than CN. 
The black circles are the solution for $\Delta t=0.5$ with CN. The simulation is still stable, although not very accurate.
On the other hand, when using RK4 even with smaller time steps ($\Delta t=0.05$ for black line and $\Delta t=0.1$ for green line) 
the simulation is unstable after only a few electron plasma periods (in all the Figures the time is in units of $f_{pe}^{-1}=2\pi/\omega_{pe}$).
For this particular case, the semi-implicit scheme presented here is at least a factor of 10 more stable.

\section{Filamentation and artificial collisionality}\label{hyper}
\begin{figure}
\centering
\includegraphics[width=0.8\textwidth]{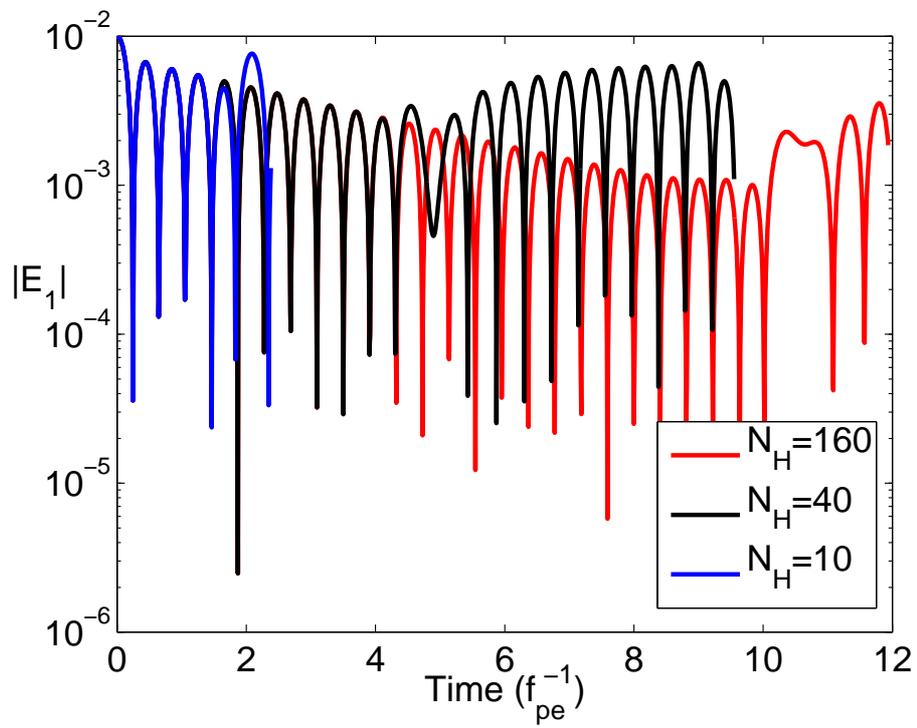}
\caption{Landau damping simulation with parameters: $\alpha_e=1$, $\varepsilon=0.01$. Recurrence phenomenon for the FH method due to filamentation. The plot shows the amplitude of $E_1$ 
(the mode in the electric field perturbed at $t=0$).
$N_H$ is the number of Hermite modes employed: $N_H=10$ (blue line); $N_H=40$ (black line); $N_H=160$ (Red line). The recurrence time is increased by increasing $N_H$, scaling
approximately as $N_H^{1/2}$.}\label{fig:recurrence}
\end{figure}

\begin{figure}
\centering
\includegraphics[width=0.8\textwidth]{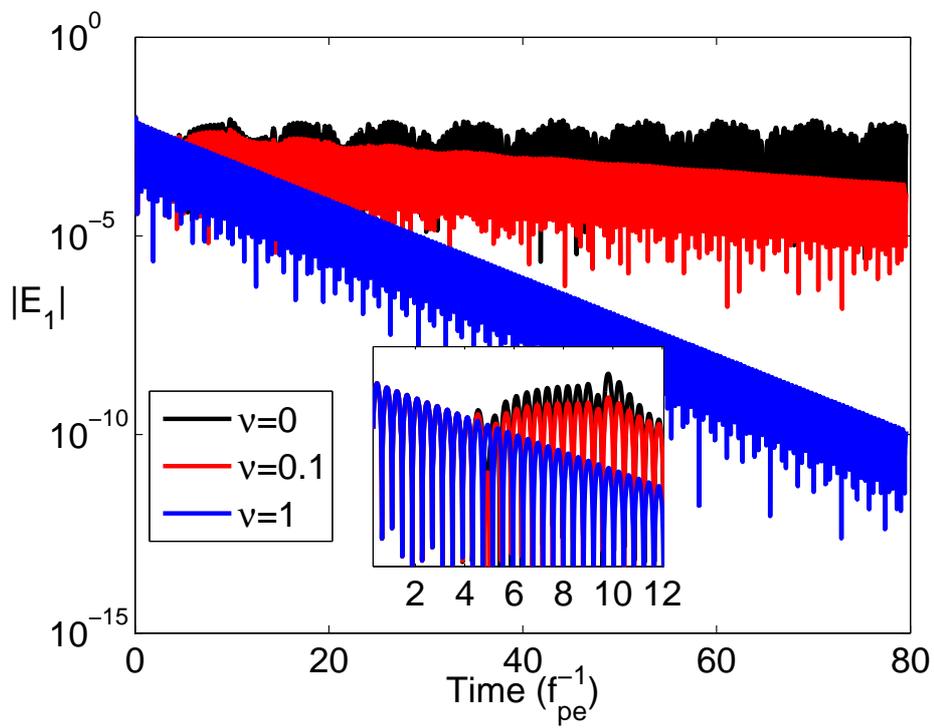}
\caption{Effect of the numerical collision term in the Landau damping simulation for the FH method, with parameters: $\alpha_e=1$, $\varepsilon=0.01$, $N_H=40$. 
When the simulation is completely collisionless ($\nu=0$, black line)
the recurrence effect due to filamentation does not allow long-time simulations. The inlet box shows a zoom-in at initial times.
The correct damping is recovered for $\nu=1$. }\label{fig:collisions}
\end{figure}

\begin{figure}
\centering
\includegraphics[width=0.8\textwidth]{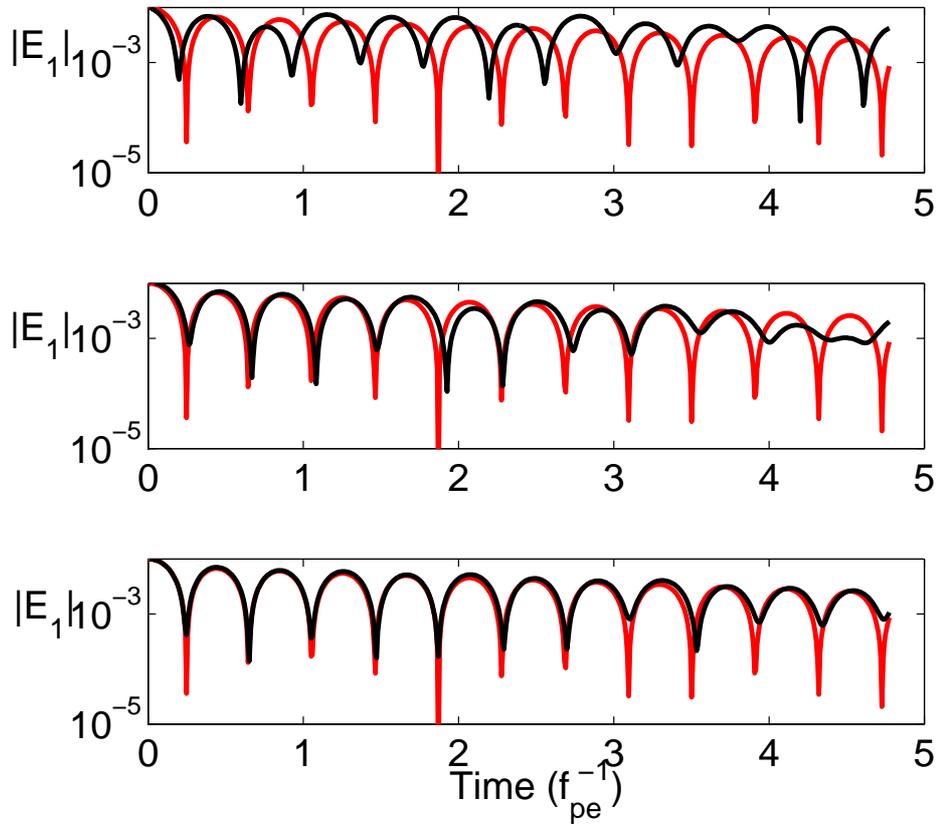}
\caption{PIC simulation of Landau damping with parameters $\alpha_e=1$, $\varepsilon=0.01$. The plot shows the amplitude of $E_1$ (the mode in the electric field perturbed at $t=0$) for different
number of particles per cell: $N_{pcel}=1600$ (top panel), $N_{pcel}=6400$ (middle panel), $N_{pcel}=25600$ (bottom panel). The black line is the PIC result and the red line is the reference solution (obtained with 
the Hermite code with $N_H=100$). }\label{fig:PIC_recurrence}
\end{figure}

\begin{figure}
\centering
\includegraphics[width=0.8\textwidth]{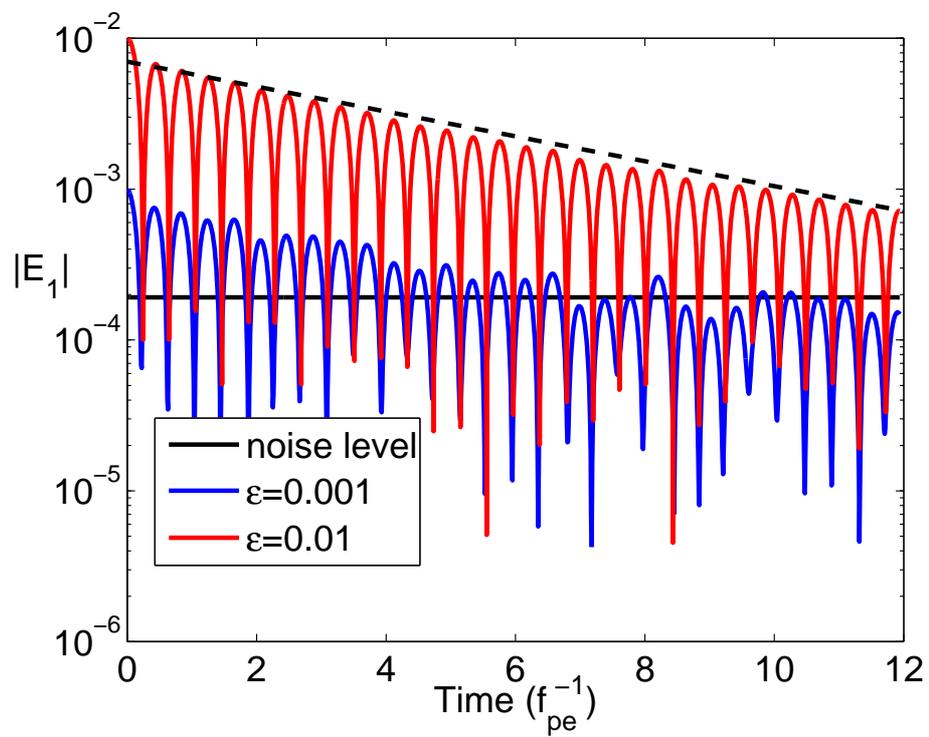}
\caption{PIC simulation of Landau damping. The number of particle per cell is $N_{pcel}=2,000,000$. Different colors denote different initial amplitudes: $\varepsilon=0.001$ (blue); 
$\varepsilon=0.01$ (red). The black solid line indicates the average noise level. The dashed line indicates the theoretical Landau damping rate.}\label{fig:PIC_vary_a}
\end{figure}

The development of increasingly smaller phase space structures in a collisionless plasma is very well known in plasma physics and 
typically referred to as the filamentation process.
A classical and well-studied example where filamentation occurs is linear Landau damping, i.e., the damping in time of an initial electric field perturbation due 
to wave-particle resonances \citep{landau46}. 
In this case filamentation is due to the presence of the factor $\exp[-ikvt]$ ($k$ is the wavevector of the perturbation) 
in the perturbed distribution function that creates oscillations with smaller
and smaller wavelengths in velocity space as time evolves.
Clearly, any discretization of the velocity space is associated with a minimum wavelength that can be resolved, and, therefore, any 
numerical simulation of the filamentation process with fixed resolution is bound to fail after a certain time.\\
In the case of the FH expansion method, it is known that the time at which the simulation is unable to capture 
filamentation due to lack of resolution in velocity space scales approximately as the square root of the number of Hermite modes, $N_H$ \citep{grant67a,canosa74,schumer98}. This time is known 
as the recurrence time: the effect of the lack of resolution is to reconstruct a distribution function similar to the initial one, from which the Landau damping starts again in a recurrent way.
Figure \ref{fig:recurrence} shows the recurrence effect for the linear Landau damping studied with the FH code (this case will be studied in more detail in Section \ref{landau}). 
Blue, black, and red lines show the amplitude of $E_1$ (the
Fourier component of the electric field perturbed at time $T=0$) for $N_H=10,40,160$, respectively. One can appreciate that the time at which the simulation manifests the 
recurrence phenomenon approximately doubles when $N_H$ is multiplied by a factor of 4, as obtained in \citet{schumer98}.\\
Several fixes have been suggested in the literature in order to overcome the filamentation process in Vlasov codes. They involve
some form of filtering or smoothing of the high order moments of the distribution function (see, e.g. \citep{klimas87, klimas94}) or, equivalently, the introduction of a weakly-collisional operator \citep{grant67a}.\\
In this work, we have tested the effect of a collisional operator. This is a purely numerical artifact, that does not represent physical collisions,
and must be used in a convergence sense, namely by tuning it such that the physical results of interest are unaffected, while the small filamentation structures are damped.
We have opted for a collision operator $\mathcal{C}$ that acts
on the coefficient $C_{n,k}$ as:
\begin{equation}
 \mathcal{C}[C_{n,k}] =-\nu \frac{n(n-1)(n-2)}{(N_H-1)(N_H-2)(N_H-3)} C_{n,k}\label{coll_operator},
\end{equation}
where $\nu$ is the collisional rate applied to the last Hermite coefficients $C_{N_H-1,k}$.
The collision operator of Eq. (\ref{coll_operator}) can be constructed as a nonlinear combination of terms involving the Lenard-Bernstein collision operator $\mathcal{C}_{LB}$ \citep{lenard58}:
\begin{equation}
 \mathcal{C}_{LB}[f] = -\nu \frac{\partial}{\partial v}\left( vf+\frac{1}{2}\frac{\partial f}{\partial v}\right),\label{lb}
\end{equation}
which, in the Fourier-Hermite space, reads:
\begin{equation}
 \mathcal{C}_{LB}[f] = -\nu n C_{n,k}.
\end{equation}
An important point is that the Lenard-Bernstein operator transformed in the
Fourier-Hermite space acts on all the coefficients $C_{n,k}$, including $n=0,1,2$, while our operator is defined in such a way that it does not change the first three Hermite modes.
The reason for choosing the form in Eq. (\ref{coll_operator}) is that it conserves charge and momentum (and energy if a fully-implicit time discretization is employed), by 
leaving $C_{0,k},C_{1,k},C_{2,k}$ unchanged.
The fact that this operator does not have a physical interpretation is not important, since our goal is to study collisionless plasmas. 
Also, the regime of validity 
of the simulation will be reduced to times for which the collisional rate is not dominant.
Of course, other forms of the collision operator might be employed: for instance \citep{parker12} proposes an iterative
version of Eq. (\ref{lb}). The crucial feature, however, is that the damping rate applied to the coefficient $C_{n,k}$ must increase (in absolute value) with increasing Hermite index $n$.\\
The effect of the artificial collision operator is to damp the highest modes of the Hermite expansion. The convergence of the series implies that 
$|C_{n+1}|/|C_n|\rightarrow 0$ for large enough $n$. However, high $n$ modes can grow due to the filamentation process or even just due to 
numerical errors. As we will show in Section \ref{twostream}, this can lead to numerical instabilities if the growth of the large $n$ modes is not suppressed artificially.\\
The effect of collisionality on the Landau damping study is presented in Figure \ref{fig:collisions}. Here $N_H=40$ and three values of $\nu$ have been used: $\nu=0, \nu=0.1, \nu=1$. 
One can notice that the correct value of Landau damping (i.e. the one obtained before recurrence when $\nu=0$) 
is recovered for a long time, when $\nu=1$ and, therefore, in this case the use of collisionality is crucial to overcome
the recurrence due to filamentation. The small box shows a zoom-in for time $T<12$.
The plot in Figure \ref{fig:collisions} must be interpreted in light of the important result presented in \citet{ng99}, and rediscussed in \citep{hilscher13}. 
It is well known that Landau damping in a collisionless plasma is due to the effect of the destructive interference of a continuous spectrum of singular eigenmodes (the Case-Van Kampen modes).
\citet{ng99} have shown that a Lenard-Bernstein collisional operator changes the spectrum of the linear Vlasov problem by replacing the singular continuous spectrum with  
a set of proper discrete eigenmodes, and that the Landau damping is recovered as a discrete mode (along with other modes). In this context,  Figure \ref{fig:collisions} clearly shows that the right 
damping rate (consistent with Landau damping) can be recovered.
One has to keep in mind, however, that although the macroscopic nature of the plasma has been preserved, the 
microscopic information associated with the high order moments of the distribution function is irreversibly modified by applying the collisional operator.
On the other hand, in most simulations of a kinetic collisionless plasma, the use of an artificial collisional operator is necessary because filamentation is an intrinsic feature.
Once again, the underlying assumption (to be verified through a convergence study) is that the use of artificial collisions  will not affect the macroscopic evolution of the system.\\
Obviously, a PIC code is not immune to filamentation problems: the use of a discrete number of super-particles implies a finite
resolution in velocity space. However, the fact that any velocity value
can be assigned to a single particle and hence that the discretized phase-space is not gridded in any standard way makes it difficult to quantify
the relationship between the number of particles and the actual resolution in velocity space. On the other hand,
it is very well-known that PIC simulations have difficulties in reproducing the fine details of a distribution function, such as high tails.
It is therefore legitimate to ask how the recurrence phenomenon shown in Figure \ref{fig:recurrence} translates to PIC simulations. 
In Figure \ref{fig:PIC_recurrence}, we show that although PIC simulations do not present recurrence similarly to Vlasov codes,
the lack of resolution in velocity space still manifests itself and produces inaccurate results. 
Top, middle and bottom panels show PIC results (black line) for number of particles per cell $N_{pcel}=1600, 6400$, and $25600$, respectively. The red line
is a reference solution calculated with the FH code (with $N_H=100$).
For all three of these cases, the correct result is lost at some point, and the solution becomes essentially noise. 
A useful way of understanding this deviation from the correct solution is to look at the signal-to-noise ratio. 
In Figure \ref{fig:PIC_vary_a}, we show the time evolution of the Landau damping test 
for different values of the initial perturbation $\varepsilon$ (see Section \ref{results} for the discussion of the 
initialization) from PIC simulations with 2,000,000 particles per cell. 
The black line indicates the noise level, which has been calculated as the maximum value of $|E_1|$ in a case without initial perturbation ($\varepsilon=0$).
Blue and red lines denote simulations with $\varepsilon=0.001$ and $\varepsilon=0.01$, respectively. Although the two simulations have the same number of particles per cell,
starting with a lower initial amplitude (blue line) (i.e. at a lower signal-to-noise level), clearly impacts the result: the Landau damping is almost immediately lost for $\varepsilon=0.001$.
In other words, the noise ground level sets the amplitude of a signal at which the simulation loses any physical interpretation.
Of course, in the Hermite method,  there is no noise ground level and the equivalent simulation (shown in Figure \ref{fig:recurrence}) 
is independent of the value of the initial perturbation (given that the value is small enough to be in the linear regime).\\
A last remark is in order with regards to collisionality and PIC codes. 
The discrete nature of the super-particles and the inevitable presence of numerical noise can in some sense be associated to the effect of a (weakly) collisional operator,
although its effect is not easily quantifiable.
The major disadvantage of PIC relative to the FH method is the bad scaling of the noise with respect to the number of particles, which, in certain applications, 
might imply the need for significant computational resources to access a true collisionless regime.
Conversely, the collision operator does not affect the performance of the FH code and a convergence study on the collision rate can be conducted quickly.

\section{Results}\label{results}
In this section we compare the computational performance of the FH spectral method 
with a standard explicit PIC code for the following four test cases: Langmuir wave, Landau damping, ion-acoustic wave and two-stream instability.
These are the same test cases discussed in Reference \citep{chen11} and routinely used to benchmark kinetic codes. 
Additionally, we present a result for an example of plasma echo. 
Each test tackles different aspects of kinetic plasma physics and the related difficulties encountered in the numerical simulations of collisionless kinetic plasmas.

In order to make the 
comparison as fair as possible, the Poisson solver in the PIC code and the FH code are identical, i.e., the Poisson equation is solved by means of a Fourier decomposition in both cases.
The PIC code employs linear interpolation, usually referred to as 'Cloud-in-Cell' (CIC) \citep{birdsall_book}. 
Note that higher order interpolation schemes have been proposed, for instance, in \citep{lewis70,evstatiev13}.
Since the focus of this work is on the discretization in velocity space, i.e., the comparison
between the spectral Hermite basis and the use of super-particles, all simulations are made for the same choice of time step and grid size.  
The comparison is characterized by the following three metrics: for each method we calculate
1) the error with respect to a 'reference' highly accurate solution as a function of CPU time and velocity discretization (number of 
Hermite modes $N_H$ for FH and number of particles per cell $N_{pcel}$ for PIC); 2) the error with respect to the 'previous' less accurate solution; 
3) the efficacy defined as the inverse of the product of CPU time and error.
The error used for all the runs is calculated as the $L_1$ norm of the difference between two solutions, averaged in time.
The error in 2) is what is actually used by a user who is performing a convergence study to decide when the solution is accurate enough.
The efficacy is a useful indicator of the cost-effectiveness of an algorithm. It measures whether an additional cost in terms of CPU time is compensated by a gain in terms of accuracy.
Clearly an algorithm performs well if the efficacy increases notably with increasing CPU time. In this regard we can anticipate that the PIC algorithm, by construction, performs badly in terms of efficacy 
since the error scales as $N_p^{-1/2}$, while the computing time scales roughly linearly with $N_p$. 
Therefore, the efficacy scales as the inverse of the square root of the CPU time, 
that is, it actually decreases with increasing
CPU time. Hence, from a pure cost-effectiveness point of view, it is \textit{never} advantageous to increase the number of particles in a PIC code to reduce the error. 
On the other hand, one is often forced to have a large number of particles such that 
the physical signal is above the noise level (see e.g. \citep{camporeale11}).\\
For all the cases discussed below, we initialize the electrons (ions) with a Maxwellian distribution function with thermal velocity $\alpha_{e}$ ($\alpha_{i}$). 
For the Langmuir wave, the Landau damping and the two-stream instability tests
the ions constitute a fixed background, while for the study of the ion-acoustic wave they evolve.
The initial electric field is initialized as:
\begin{equation}
 E(x,t=0) = \frac{L}{2\pi}\varepsilon\sin(2\pi x/L).
\end{equation}
$L$ is the box length ($L=4\pi$ for all runs), and $\varepsilon$ is the amplitude of the initial perturbation.
In Fourier space such initialization corresponds to:
\begin{equation}
 E_{-1}=E_{1}=-\varepsilon\frac{L}{4\pi}
\end{equation}
and the density is initialized consistently.
For all runs the number of Fourier modes is equal to $2N+1=33$ and the time step is $\Delta t=0.05$. 
In all runs presented here we have used the CN scheme for the X-shift of the FH code.
All the codes are written in MATLAB and run on an Intel Xeon 3.40 GHz Linux box.

\subsection{Langmuir wave}
\begin{figure}
\centering
\includegraphics[width=1\textwidth]{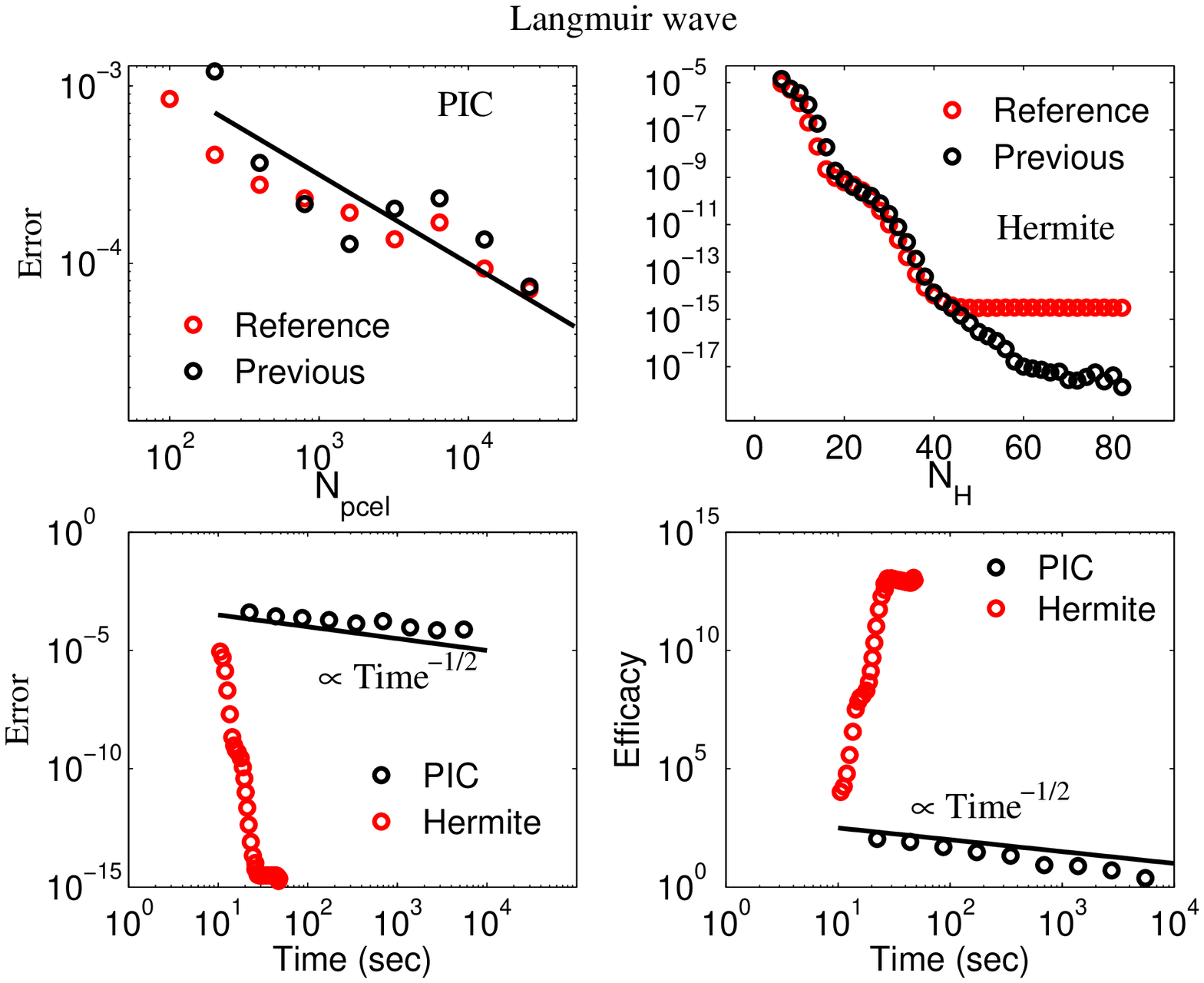}
\caption{Langmuir wave with parameters $\alpha_e=0.1$, $\varepsilon=0.01$. Top left: PIC simulation; error as a function of number of particles per cell $N_{pcel}$. Red and black circles represent the error
calculated with respect to a reference solution (with $N_{pcel}=102400$) and previous less accurate solution, respectively. The black solid line indicates the scaling $N_{pcel}^{-1/2}$.
Top right: Hermite simulation; error as a function of number of Hermite modes $N_H$. Red and black circles represent the error
calculated with respect to a reference solution (with $N_H=100$) and previous less accurate solution, respectively.
Bottom left: error as a function of CPU time (in seconds); black circles for PIC, red circles for Hermite. Bottom right: efficacy as a function of CPU time (in seconds); black circles for PIC, red circles for Hermite.
}\label{fig:langmuir}
\end{figure}

The parameters are chosen as follows: $\alpha_e=0.1$, $\varepsilon=0.01$. In this case, the electrons are relatively cold and therefore are not subject to Landau damping. 
The initial perturbation causes the electrons to oscillate collectively around the fixed ion background.
Figure \ref{fig:langmuir} shows the error achieved with respect to the reference solution (red circles) and the previous solution (black circles) for PIC (top-left panel) and FH codes (top-right panel), 
as a function of the resolution in velocity space.
The theoretical scaling $N_{pcel}^{-1/2}$ (black line) is clearly recovered for PIC. 
The difference in errors are huge, such that the most accurate solution obtained with PIC (102,400 particles per cell) is still less accurate than
the solution obtained with only 4 Hermite modes. One can also notice that for the FH code the error with respect to the reference solution flattens for $N_H>40$.
The way this result is reflected in terms of CPU time is shown in the bottom-left panel of Figure \ref{fig:langmuir}. Here the CPU time is reported on the horizontal axis in seconds, 
and the error (with respect to the reference solution) is on the vertical axis.
The black circles are PIC results, while the red circles are FH results. Strikingly, the most accurate solution for the FH code and the least accurate solution for the PIC code
are obtained with roughly the same CPU time!
The accuracy of these solutions differs by about 10 orders of magnitude.
Finally, the efficacy is plotted on the bottom-right panel of Figure \ref{fig:langmuir} (black circles are PIC results, red circles are FH results). 
As anticipated the PIC efficacy decreases by increasing CPU time, and is consistent with the theoretical prediction.
On the other hand, the FH efficacy increases by 7 orders of magnitude when the CPU time increases by less than a factor of 10.

\subsection{Landau damping}\label{landau}
\begin{figure}
\centering
\includegraphics[width=1\textwidth]{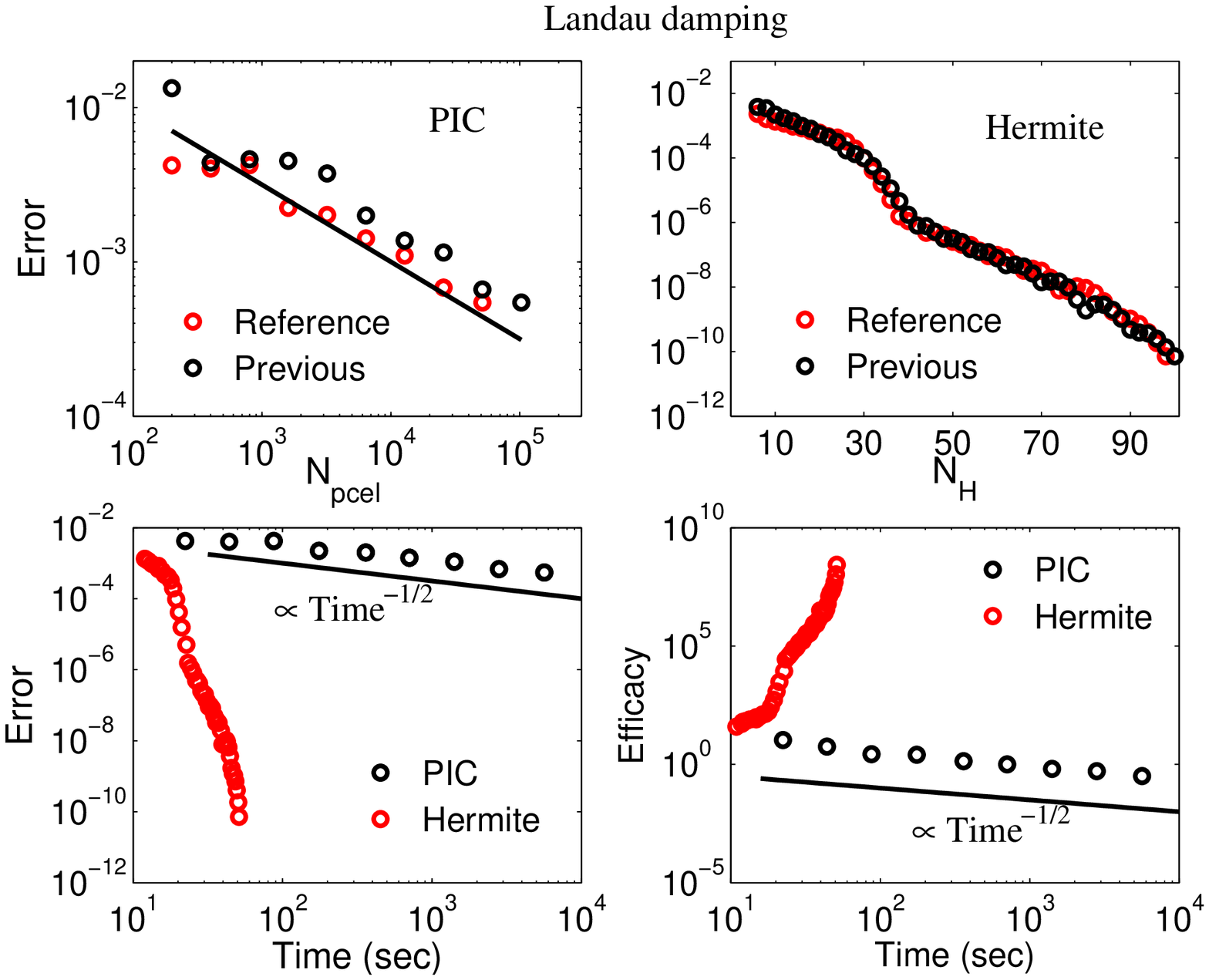}
\caption{Landau damping with parameters $\alpha_e=1$, $\varepsilon=0.01$. Top left: PIC simulation; error as a function of number of particles per cell $N_{pcel}$. Red and black circles represent the error
calculated with respect to a reference solution (with $N_{pcel}=102400$) and previous less accurate solution, respectively. The black solid line indicates the scaling $N_{pcel}^{-1/2}$.
Top right: Hermite simulation; error as a function of number of Hermite modes $N_H$. Red and black circles represent the error
calculated with respect to a reference solution (with $N_H=100$) and previous less accurate solution, respectively.
Bottom left: error as a function of CPU time (in seconds); black circles for PIC, red circles for Hermite. Bottom right: efficacy as a function of CPU time (in seconds); black circles for PIC, red circles for Hermite.
}\label{fig:landau}
\end{figure}

Landau damping is a classical kinetic effect in warm plasmas. It is due to the effect of particles in resonance with an initial wave perturbation that results in the damping of the wave \citep{brambilla_book}.
The parameters are chosen as follows: $\alpha_e=1$, $\varepsilon=0.01$.
Landau damping, as discussed in Section \ref{hyper}, is a particularly challenging problem for kinetic codes, because of the continuous filamentation 
in velocity space. For the FH code, this translates to the fact that more Hermite modes are needed as the simulation evolves, with respect to the Langmuir wave case.
Figure \ref{fig:landau} summarizes the results in the same format of Figure \ref{fig:langmuir}.
The top left and right panels show the convergence study of error as a function of $N_{pcel}$ and $N_H$ for the PIC and FH codes, respectively.
The bottom left panel shows the error as a function of CPU time, and the bottom right panel represents the efficacy as a function of CPU time. 
Similarly to the Langmuir wave case, the FH code greatly outperforms the PIC code: the efficacy for the FH method increases by 
about 6 orders of magnitude when the CPU time increases only by a factor of 6, in the interval from 10 to 60 seconds. 
Conversely, the difference in error between the two methods is as large as 8 orders of magnitude, for equal CPU time.

\subsection{Ion-acoustic wave}\label{ionacoustic}
\begin{figure}
\centering
\includegraphics[width=1\textwidth]{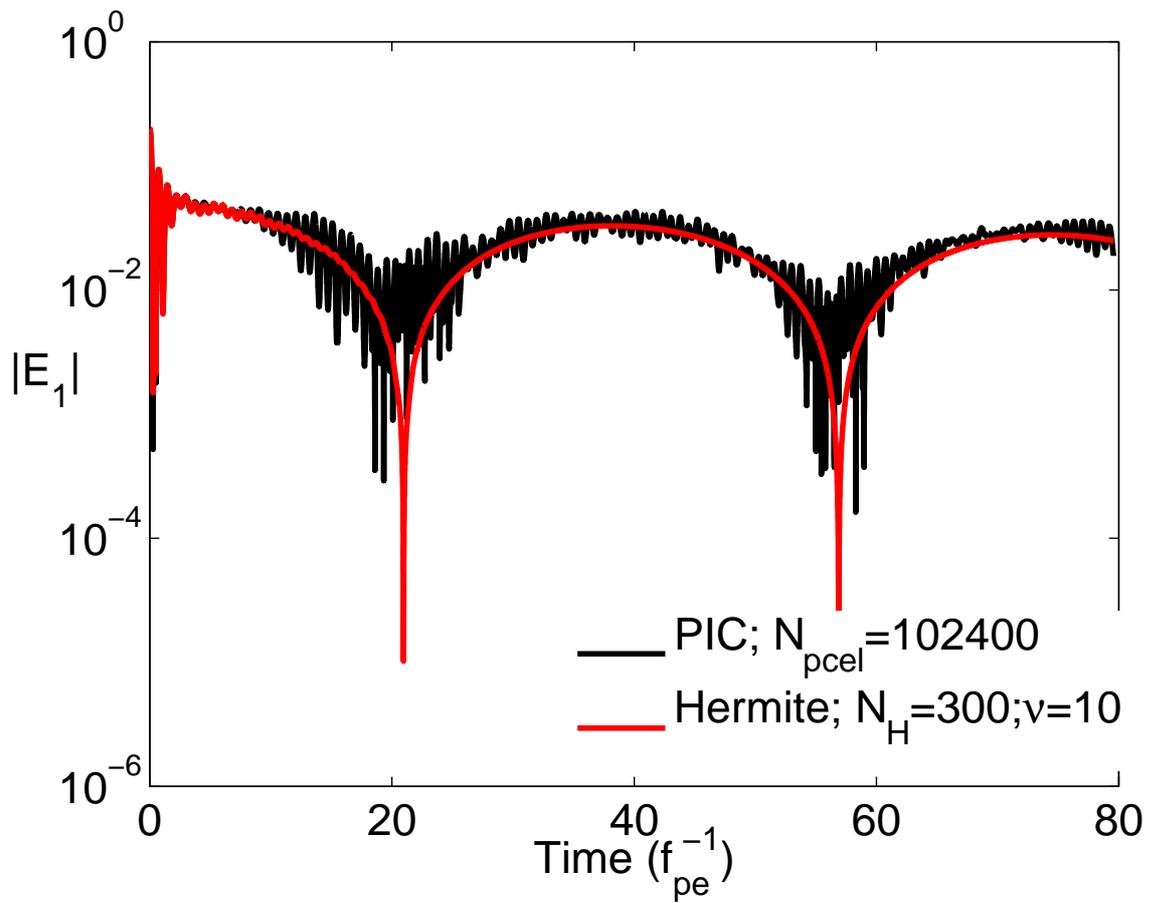}
\caption{Time evolution of the amplitude of perturbed component of the electric field for the ion acoustic wave. Black and red lines are 
for PIC and FH results, respectively. PIC is run with $N_pcel=102400$ and FH with $N_H=300$ and $\nu=10$.}\label{fig:ion_acoustic_time}
\end{figure}

\begin{figure}
\centering
\includegraphics[width=1\textwidth]{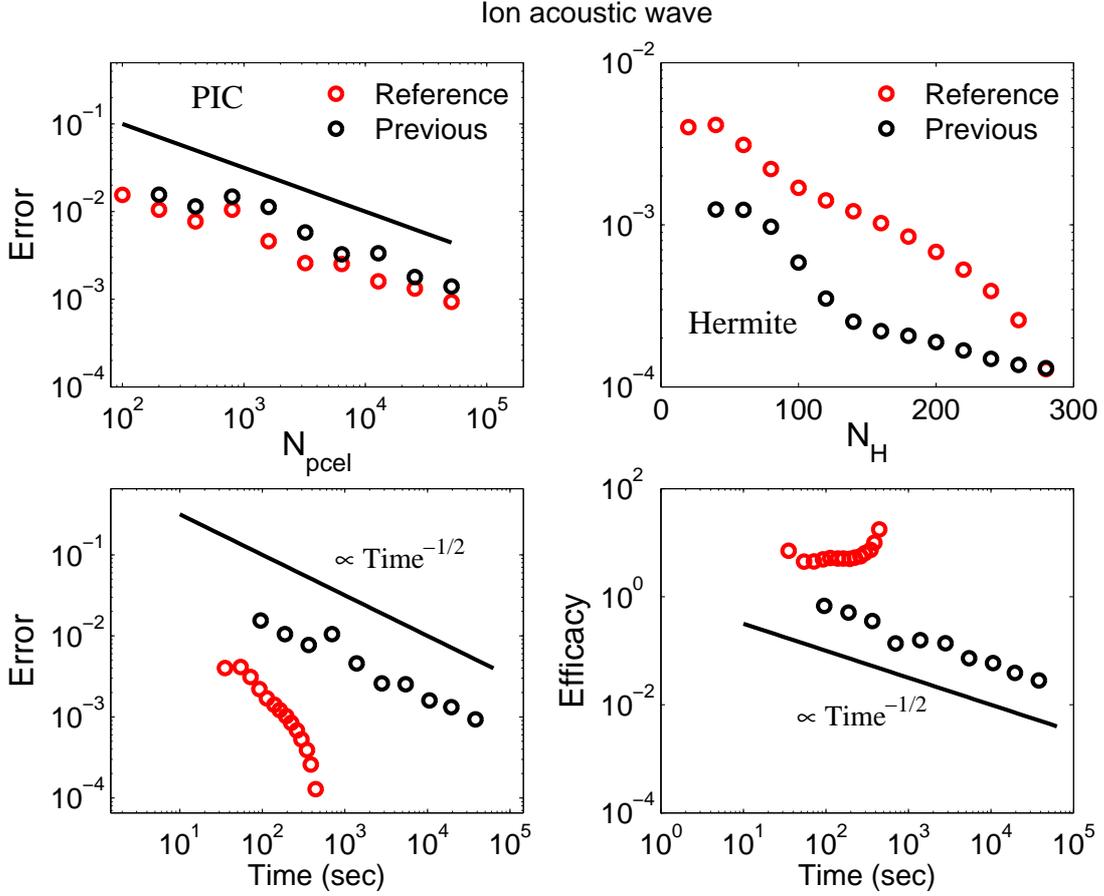}
\caption{Ion acoustic wave with parameters $\alpha_e=1$, $\alpha_i=0.0074$, $\varepsilon=0.2$. Top left: PIC simulation; error as a function of number of particles per cell $N_{pcel}$. Red and black circles represent the error
calculated with respect to a reference solution (with $N_{pcel}=102400$) and previous less accurate solution, respectively. The black solid line indicates the scaling $N_{pcel}^{-1/2}$.
Top right: Hermite simulation; error as a function of number of Hermite modes $N_H$. Red and black circles represent the error
calculated with respect to a reference solution (with $N_H=300$) and previous less accurate solution, respectively.
Bottom left: error as a function of CPU time (in seconds); black circles for PIC, red circles for Hermite. 
Bottom right: efficacy as a function of CPU time (in seconds); black circles for PIC, red circles for Hermite.}\label{fig:ion_acoustic}
\end{figure}

The ion-acoustic wave is a typical example of a multi-scale phenomenon in kinetic plasma physics, since it requires both electrons and ions dynamics.
Hence, the distribution function of both species is evolved. The ion acoustic wave is initialized by perturbing the ion population only.
The mass ratio between the two species is equal to 1836, and the ratio of electron and ion temperature is $10$.
This results in a ratio of electron to ion thermal velocities of 
$\alpha_{e}/\alpha_{i} = 135$. The initial perturbation is $\varepsilon=0.2$. Hence, the simulation is performed in a nonlinear regime.
For this case, we have used a collisionality with $\nu=10$.\\
Figure \ref{fig:ion_acoustic_time} shows the time evolution of $|E_1|$ for PIC (black line, $N_{pcel}=102400$) and FH (red line, $N_H=300$, $\nu=10$). The results are in good agreement,
although PIC results are still very noisy even with such a large number of particles per cell.
Figure \ref{fig:ion_acoustic} shows the errors and the efficacy with the same format as before.
Once again, the scaling $N_{pcel}^{-1/2}$ is approximately recovered for PIC, and the performance of the two methods differ
by orders of magnitude. For instance, to obtain an error of the order of $10^{-3}$ the PIC code takes about 3 orders of magnitude
more CPU time than FH.

\subsection{Two-stream instability}\label{twostream}
\begin{figure}
\centering
\includegraphics[width=0.8\textwidth]{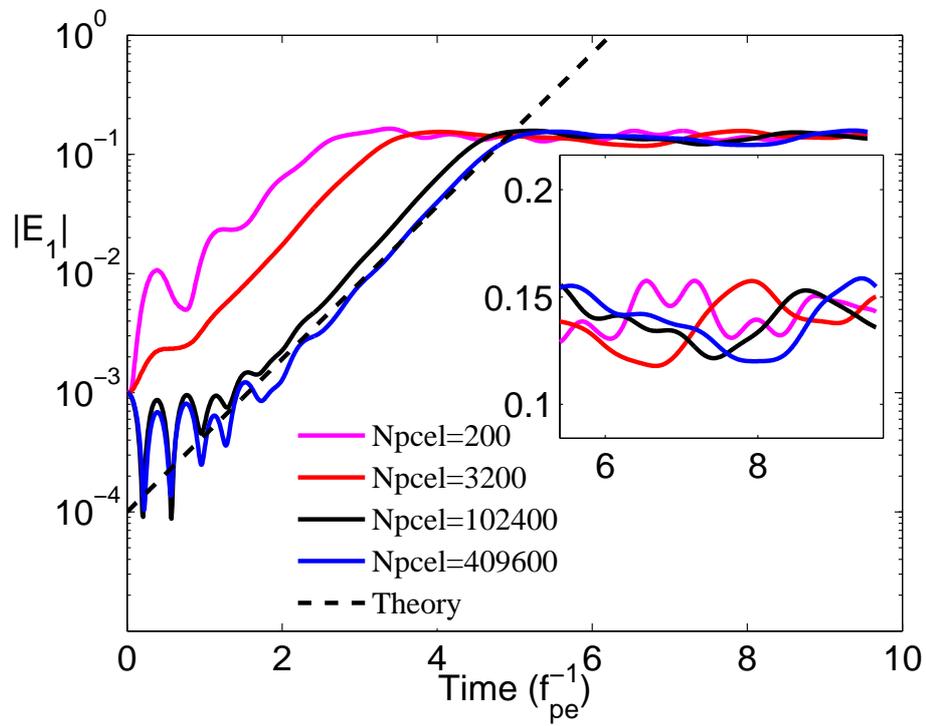}
\caption{Time evolution for the two-stream instability from PIC simulations, with different number of particles per cell ($\alpha_e=0.5$, $\varepsilon=0.001$, $U=1$). $N_{pcel}=200$ (magenta); $N_{pcel}=3200$ (red); $N_{pcel}=102400$ (black); $N_{pcel}=409600$ (blue).
The dashed line represents the theoretical linear growth rate. The small box shows a zoom-in of the post-saturation phase.}\label{fig:twostream_PIC}
\end{figure}

\begin{figure}
\centering
\includegraphics[width=0.8\textwidth]{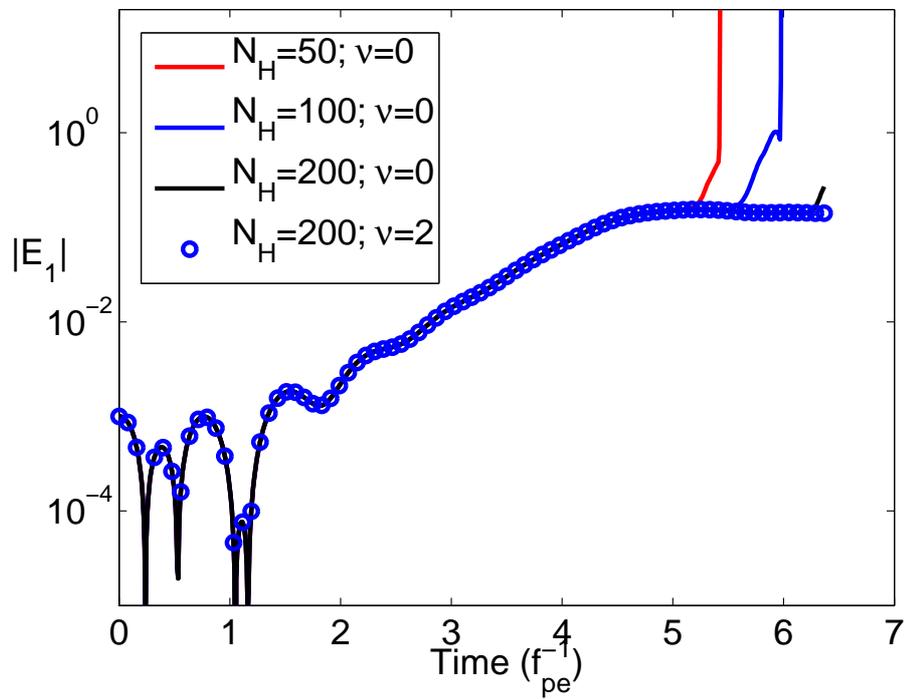}
\caption{Two stream instability with the FH method ($\alpha_e=0.5$, $\varepsilon=0.001$, $U=1$). Red, blue, and black lines denote runs without collisions ($\nu=0$) for $N_H=50,100,200$, respectively. 
Blue circles show the solution for $N_H=200$ and $\nu=2$.}\label{fig:twostream_hermite}
\end{figure}

\begin{figure}
\centering
\includegraphics[width=1\textwidth]{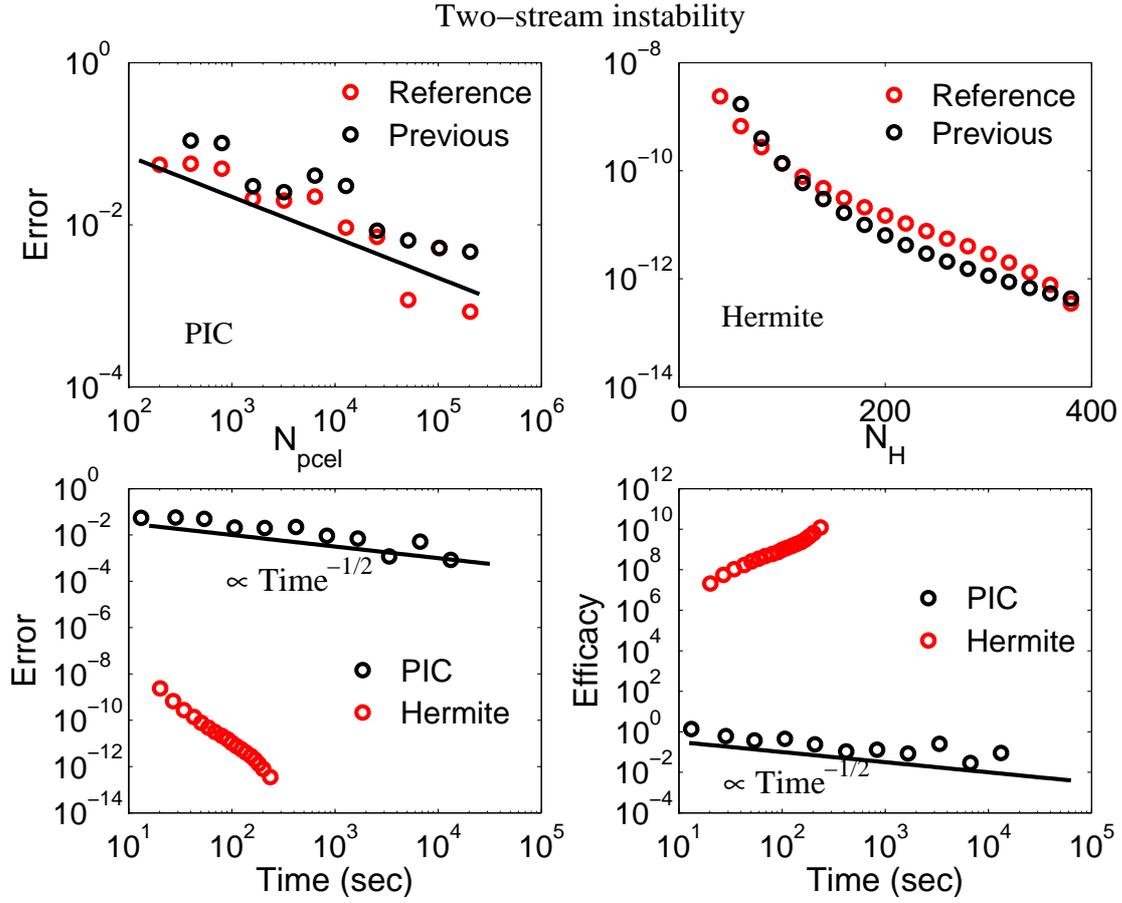}
\caption{Two-stream instability with parameters $\alpha_e=0.5$, $\varepsilon=0.001$, $U=1$. Top left: PIC simulation; error as a function of number of particles per cell $N_{pcel}$. Red and black circles represent the error
calculated with respect to a reference solution (with $N_{pcel}=819,200$) and previous less accurate solution, respectively. The black solid line indicates the scaling $N_{pcel}^{-1/2}$.
Top right: Hermite simulation; error as a function of number of Hermite modes $N_H$. Red and black circles represent the error
calculated with respect to a reference solution (with $N_H=400$) and previous less accurate solution, respectively.
Bottom left: error as a function of CPU time (in seconds); black circles for PIC, red circles for Hermite. 
Bottom right: efficacy as a function of CPU time (in seconds); black circles for PIC, red circles for Hermite.}\label{fig:twostream}
\end{figure}

The two stream instability is excited when the distribution function of a species is formed by two populations streaming in 
opposite directions with a large enough relative drift velocity. We initialize the electron distribution function as
two counter-streaming Maxwellians with equal temperature:
\begin{equation}
 f=0.5e^{\left(\frac{v-U}{\alpha_{e}}\right)}+0.5e^{\left(\frac{v+U}{\alpha_{e}}\right)}\label{drift-maxw}
\end{equation}
where $U$ is the drift velocity.\\
In principle, a distribution function of the form (\ref{drift-maxw}) would require a large number of Hermite polynomials in order to be described accurately
by the expansion in Eq. (\ref{f}).  However, a way to overcome this difficulty is to split the distribution function in (\ref{drift-maxw}) 
into two distinct populations corresponding to each drifting Maxwellian, 
and to solve Eq. (\ref{main_eq}) separately for each population. As already anticipated, including a free parameter $u$
in the definition of the argument $\xi$ allows the description of a drifting Maxwellian by means of only one FH coefficient in Eq.(\ref{f}) (by simply choosing $u_e=\pm U$).\\
For this test case, we have chosen the following parameters: $\alpha_e=0.5; U=1; \varepsilon=0.001$.\\
Figure \ref{fig:twostream_PIC} shows the evolution in time of $|E_1|$ obtained by PIC simulations with a different number of particles per cell (magenta=200; red=3,200; black=102,400; blue=409,600).
The black dashed line represents the theoretical growth rate of the instability, calculated by linear theory.
One can notice that the convergence of the PIC simulation is very slow, particularly due to the fact that the initial transient that gives rise to the instability 
is very dependent on the number of particles, i.e. the noise level. Also, although each run saturates at approximately the same level, the saturation time is different, making the post-saturation
phase not comparable between different runs (the smaller box in Figure \ref{fig:twostream_PIC} shows a zoom-in of the post-saturation dynamics, in linear scale).
For this reason, the errors used in this section have been calculated only in the linear regime region, i.e. for times smaller than 5.\\
Interestingly, capturing fine structures in the distribution function (in velocity space) becomes important, in the two-stream instability, only in the post-saturation phase.
This can be clearly appreciated in Figure \ref{fig:twostream_hermite}. Here, we plot the time evolution of $E_1$, for FH simulations with different values of $N_H$.
Even with a low number of Hermite modes ($N_H=50$, red line), the linear stage of the instability is well captured. However, the simulation becomes numerically unstable 
in the post-saturation phase. By increasing the number of Hermite modes ($N_H=100$, blue line; $N_H=200$, black line), the time at which the 
simulation becomes unstable is postponed. This indicates that a large number of Hermite modes is needed because fine structures in the velocity space continue 
to form after the instability has saturated, i.e. the plasma is developing filamentation. The solution indicated with circles is obtained with $N_H=200$ using a collisional operator with $\nu=2$.
The numerical collisions do not change the result macroscopically, but allow longer simulations without the need of a prohibitively large number of Hermite modes.
The fact that the FH simulation becomes numerically unstable if collisions are not applied may be viewed as a shortcoming of the method.
On the contrary, we regard this feature as beneficial, especially when compared with the PIC code. In fact, the appearance of the numerical instability is a clear signal
of the lack of resolution in velocity space. The straightforward approach to overcoming this problem is by increasing $N_H$ and/or introducing artificial collisions. On the other hand,
in the PIC simulation the only way to understand that the results are not converged is to perform an expensive convergence study. In other words,
the PIC simulation will keep running, potentially for a long time, wasting computational resources, and producing results that have no physical meaning.\\
Figure \ref{fig:twostream} presents the errors and efficacy for the two-stream instability case. The results are qualitatively similar to the previous cases, once again highlighting
the high performance of the FH method with respect to PIC.

\subsection{Plasma echo}
\begin{figure}
\centering
\includegraphics[width=1\textwidth]{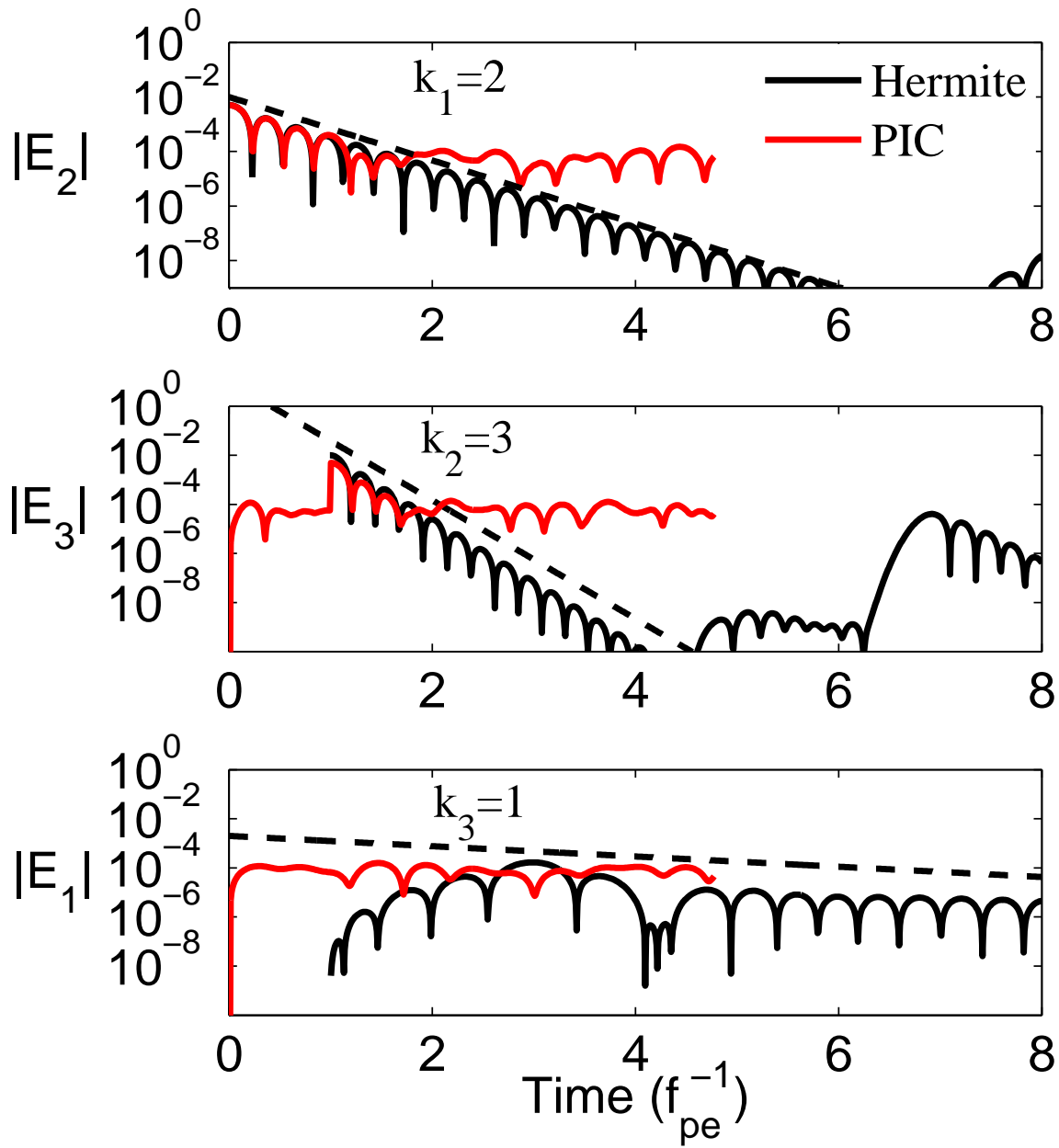}
\caption{Plasma echo. Time evolution of $|E_2|$ (top), $|E_3|$ (middle), and $|E_1|$ (bottom). Black line is for Hermite ($N_H=400$), red line is for PIC ($N_{pcel}=2,000,000$).
$E_2$ is excited at $T=0$ and $E_3$ is excited at $T=1$. The resulting echo at $T=3$ manifests in the excitation of $E_1$.
}\label{fig:echo}
\end{figure}

In contrast to the previous tests, the rationale for this case is not to compare results between PIC and FH methods but to provide an example 
of a fundamental kinetic plasma phenomenon that is very hard  to capture with PIC, but is still straightforward to simulate with the FH code.\\
The basic theory behind plasma echo has been studied in the seminal paper by \citet{gould67}, and investigated experimentally by \citet{malmberg68}.
Since then the theory has been revised and rediscussed by several authors \citep[e.g.][]{brackbill72,best74,porkolab78,hou11,mouhot11,pezzi13}.
Plasma echo has been suggested as the 'archetype' of numerical tests for one-dimensional Vlasov codes in \citep{galeotti06}.
The plasma echo mechanism is briefly explained in \citep{porkolab78} as the following. 
An initial electric field perturbation with wavevector $k_1$ is initialized at time $t_1=0$ in a stable Maxwellian plasma. 
This perturbation damps according to Landau damping and modulates the distribution function through a term of the form $\exp(-ik_1x+ik_1vt)$.
At time $t_2>t_1$ a second perturbation is initialized, with wavevector $k_2$, which also damps according to
its linear damping rate. This second perturbation will act on the distribution function via a first order term of the form $\exp[ik_2x-ik_2v(t-t_2)]$.
However, a second order contribution will arise from the interaction between the two perturbations, with the form $\exp[i(k_2-k_1)x+ik_2vt_2-i(k_2-k_1)vt]$.
While the first order terms will damp in time because of phase-mixing, the second order term loses its $v$ dependence at time $t_3=k_2t_2/(k_2-k_1)$, and therefore will not phase-mix around that time.
As a consequence, a new perturbation with wavevector $k_3=k_2-k_1$ will appear, peaking at time $t_3$, and successively causing Landau damping.
The plasma echo is a nonlinear effect due to the interaction of two small amplitude damping waves. Interestingly, the time $t_3$ can be made much larger than $t_1$ and $t_2$ such that
the reformation of the electric field after the two parent perturbations have long been damped can effectively appear as an 'echo'.\\
The simulation is initialized with a perturbation $\varepsilon=0.01$ in $E_2$. At time $T=1$, a second perturbation
is generated, again with amplitude $\varepsilon=0.01$ but in $E_3$.
The resulting echo appears at time $T=3$ for the Fourier mode $E_1$.
Figure \ref{fig:echo} shows the amplitudes of $E_2$, $E_3$, and $E_1$ respectively in the top, middle and bottom panels.
Black solid lines are for the FH code, and red lines are for PIC. The dashed black line indicates the theoretical
Landau damping of each mode.
Clearly, the FH results are in excellent agreement with the theoretical prediction discussed above, showing the peaking of
$E_1$ at time $T=3$.
We note that we have used $\nu=0$ and $N_H=400$ in this example and some recurrence can be seen in the FH code for $T>4$
The PIC simulations are run with $N_{pcel}=2,000,000$. This is of course an extremely large number of particles
per cell, larger than what is routinely used. However, the PIC simulation completely misses the echo phenomenon, simply because a noise ground level comparable
with the echo signal is generated for $E_1$ since the beginning of the simulation (bottom panel), and both $E_2$ and $E_3$ have reached the noise ground level, by the time the echo should be generated. \\
It is important to emphasize that the plasma echo is not merely an academic exercise.
The physics of wave-wave interactions for small amplitude perturbations constitutes the building block of the weak-turbulence theory, important, for instance, for magnetized plasmas in the solar wind.

\section{Conclusions}
The focus of this work was to compare the discretization of the velocity space in the Vlasov-Poisson system based on the Hermite expansion with the use of super-particles adopted in the standard PIC method.

We have described the spectral method based on the FH expansion previously discussed in \citet{schumer98}. We have introduced two important improvements with respect to the work in \citep{schumer98}. First, we have employed a semi-implicit time discretization. Our scheme is still second order accurate in time, but we have shown that it is numerically more stable than the explicit scheme previously used. Second, we have discussed the use of an artificial collisional term, whose scope is to damp high order Hermite modes. The collision term has a double effect: it prevents the recurrence effect in perturbations which are damping, and avoids the growth of higher order modes that can lead to numerical instabilities.
As such, it is an essential ingredient of the Hermite spectral method as it takes care of the unavoidable filamentation problem typical of collisionless plasmas.\\
The comparisons between the FH method and PIC have been carried out from the perspective of computational efficiency and efficacy. The efficiency is given by the error achieved for a given CPU time, while the efficacy is a measure of the cost-effectiveness of the algorithm and is defined as the inverse of the product of the error and the CPU time.
We have studied the following standard kinetic plasma test cases: Langmuir waves, Landau damping, ion-acoustic wave, two-stream instability. For each test case, we have evaluated the error of the solutions obtained with the two methods, and the CPU time for each run. In all these examples, the FH method obtains a much accurate solution than PIC, at a fraction of the computational cost.
For instance, for the two-stream instability case, the FH method produces a result that is about 10 order of magnitudes more accurate in about 1/50 of the CPU time taken by PIC. We have also studied an example of the plasma echo phenomenon. For this case, we have shown that the standard PIC method cannot capture the nonlinear wave-wave interactions underlying the echo physics, despite using an astonishing 2,000,000 particles per cell.\\
It has to be emphasized that whilst the PIC method has been investigated and developed for the last several decades, methods based on a spectral expansion of the distribution function (either Hermite or Fourier) are much less mature. For this reason, in our comparisons we have not taken into account any strategy that could reduce the PIC noise. Notably, methods based on $\delta$f or re-mapping of the distribution function have been successful to reduce particle noise, and hence enhance the accuracy of PIC simulations (see, e.g. \citep{chehab05, wang11, deng14}).
Of course, this aspect lies in the realm of algorithm optimization. For the sake of a fair comparison, however, we reckon that it is preferable to deal with the unoptimized method, as this exposes the real limitations of each algorithm. Furthermore, for what concerns PIC, there are many different ways to reduce the noise. This opens the possibility of trying and comparing several different algorithms, each with its own pros and cons, and it is likely that different test cases will be more or less favorable for a given optimization.
On the other hand, the optimization of the
Fourier-Hermite method has practically never been investigated, although one could envision a strategy where the Hermite basis changes in time in order to use the least number of modes. 
Another consideration is that the PIC method has, by construction, an unoptimal scaling of the error with the CPU time: its efficacy decreases with increasing CPU time. 
Denoising techniques can be thought as an attempt to counteract such bad scaling. Of course, this might be beneficial for practical applications, but should not change the overall scaling of 
the method with the number of particles. It is also not obvious whether such techniques will actually increase the PIC performance from an efficacy perspective.\\
As a final comment, we want to remark some of the limitations and challenges of the FH method relative to PIC. In general, while the benchmark tests studied here are standard in the PIC community, near-equilibrium Maxwellian plasmas favor FH over PIC (since the Hermite basis naturally links to Maxwellians). On the other hand, FH is not well suited for cold plasmas/beams, which are instead treated very effectively by PIC. A different spectral basis could be sought for these cases. Similarly, if the plasma develops very complicated structures in the distribution function, many Hermite modes might be necessary to achieve convergence. This can become a serious a problem in multidimensional applications, unless some form of basis optimization is successfully developed. This is likely the most critical need of FH for application to collisionless plasmas. We have also observed that numerical instabilities can arise in the FH method for long wavelengths and large initial perturbations. We have verified that a 
possible fix/mitigation of this problem is to introduce a k-dependent collision operator (a detailed analysis of this approach will be presented elsewhere). 
On the contrary, the PIC algorithm is very robust.\\
In conclusion, considering the computational challenges associated with multi-scale plasma physics and the upcoming exascale computing era, it is obvious that numerical methods that can use efficiently the ever-increasing available computational power are critical. Despite the proof-of-principle nature of our study, our results indicate that a method based on a spectral expansion of the velocity part of the distribution function should be given serious consideration for this task. In addition to potential orders of magnitude higher efficacy (relative to PIC), we believe that another major strength of the Hermite spectral approach is its natural ability to bridge between regions where the plasma behaves like a
fluid (hence it is characterized by a low number of moments) and regions where the microscopic/kinetic physics is important (thus requiring a higher number of moments). Strategies that can accelerate the convergence of the Hermite basis will be critical for the success of the FH method in multidimensions, and we hope that this paper can stimulate some research in this area.

\section{Acknowledgement}
The authors would like to thank L. Chacon, W. Daughton, C. Huang, V. Roytershteyn, X. Tang for useful conversations.
This work was funded by the Laboratory Directed Research
and Development program (LDRD), U.S. Department of Energy Office of
Science, Office of Fusion Energy Sciences, under the auspices of the
National Nuclear Security Administration of the U.S. Department of
Energy by Los Alamos National Laboratory, operated by Los Alamos
National Security LLC under contract DE-AC52-06NA25396.





\bibliographystyle{plainnat}






\newpage

\clearpage

\clearpage

\end{document}